\documentclass[prb,aps,twocolumn,showpacs,amsmath,amssymb,floatfix,
citeautoscript]{revtex4}
\usepackage{graphicx}
\usepackage{amsmath}
\usepackage{amssymb}
\usepackage{times}

\newcommand{\Tk}{T_{\rm K}}

\begin{document}

\title{Conductance of Quantum Impurity Models from Quantum Monte Carlo}

\author{Dong E. Liu, Shailesh Chandrasekharan, and Harold U. Baranger}
\affiliation{
Department of Physics, Duke University, Box 90305, Durham, North Carolina
27708-0305 USA
}
%\date{\today}
\date{July 29, 2010}

\begin{abstract}
% The conductance of two quantum impurity models is calculated using a 
% world-line quantum Monte Carlo (QMC) method. The two Anderson impurity
% models, 
% one with two-fold and another with four-fold degeneracy, represent two types 
% of quantum dots. 
The conductance of two Anderson impurity models, one with two-fold and another
with four-fold degeneracy, representing two types of quantum dots, is calculated
using a world-line quantum Monte Carlo (QMC) method.
Extrapolation of the imaginary time QMC data to zero frequency yields the linear
conductance, which is then compared to numerical renormalization group results
in order to assess its accuracy. We find that the method gives excellent results at low
temperature ($T\alt \Tk$) throughout the mixed valence and Kondo regimes, but it
is unreliable for higher temperature. 
\end{abstract}


\maketitle

Quantum dots provide a highly controlled and tunable way to study a range of
quantum many-body physics: various quantum impurity models and their associated
Kondo effects \cite{GoldhaberGRev07, Jarillo05, MakarovskiPRB07,
MakarovskiPRL07, Potok06, Roch09},
%[single level Anderson impurity, two-channel Kondo, underscreened Kondo, 
% and SU(4) Kondo, for instance]
tunneling with dissipation \cite{BomzePRB09}, and Luttinger liquid effects
\cite{WesterveltScience96,AshooriPRL99}, to name a few. The crucial experimental
observable in these situations is the conductance; thus, calculating the
conductance is a key task for both analytic and numerical approaches. Numerical
methods have indeed been developed \cite{WilsonRMP75,NRGreviewRMP08, Izumida98,
Costi01}, with remarkable agreement for small systems between theory and 
experiment \cite{OliveiraEPL09}. But these methods scale poorly for the larger,
more complex multi-dot systems \cite{LindelofFujisawa08,Sachrajda09} that are
currently of great interest.
%  (>2 qdots, 2 CNT dots). 
Here we implement and test a way to calculate the conductance from a
path-integral quantum Monte Carlo (QMC) calculation. While it yields less
information than NRG in simple systems 
(e.g.\ a single quantum dot), the method should scale readily to more
complicated systems. Results for two Anderson-type impurity models show that the
method works very well at low temperature.

% (i) single Anderson impurity and (ii) an SU(4) Anderson impurity (CNT). 
% For application to four quantum dots see Ref.\,\onlinecite{Dong_4qdots}.  

For calculations of the conductance in simple quantum dot systems, the most
accurate results are obtained using the numerical renormalization group (NRG)
method \cite{WilsonRMP75,NRGreviewRMP08, Izumida98, Costi01}. NRG becomes slow
and even impractical, however, if there are many leads, a many-fold degeneracy,
or more than a few interacting sites. In such situations, the world-line quantum
Monte Carlo (QMC) method could be a valuable alternative since it scales nicely
as the problem size increases. However, QMC is formulated in imaginary time
rather than real time: to extract dynamic properties one must transform from
imaginary back to real time. The statistical error in the QMC data makes this an
ill-posed problem, for which various extrapolation and continuation methods have
been developed \cite{JarrellPhysRep96}. To obtain the conductance of interest
here, we extrapolate to zero frequency the appropriate correlation function
evaluated using QMC at the imaginary Matsubara frequencies
\cite{IzumidaQMC99,Louis03,BokesPRB04,BokesJCP09,Syljuasen07}. This has been
used, for instance, to study a one-dimensional Hubbard chain coupled to
non-interacting leads in the absence of the Kondo effect \cite{Syljuasen07}. 

The aim of this paper is to test the validity of the extrapolation method for
Anderson impurity models in both the mixed valence and Kondo regimes. We study
the linear conductance using QMC in two models: a single impurity Anderson model
with either two-fold or four-fold degeneracy. The standard two-fold degenerate
model is a simplified representation of a single GaAs quantum dot connected to
leads \cite{GoldhaberGRev07}. The four-fold degenerate model represents a
quantum dot in a carbon nanotube in which there is an additional orbital
degeneracy from the helicity of the states \cite{GoldhaberGRev07, Jarillo05,
MakarovskiPRB07, MakarovskiPRL07}. This orbital degeneracy is present in both
the discrete states in the dot and the extended states in the carbon nanotube
leads.

Consider a model, then, in which a single level with Coulomb repulsion $U$
represents the quantum dot (which we also refer to as the impurity site)
and is coupled to two non-interacting bands, left ($L$) and right ($R$). The
degeneracy of both the discrete level and the free electrons is $M$; we will
consider the two cases $M \!=\! 2$ (standard single-level Anderson model) and $M
\!=\! 4$ (both spin and orbital degeneracy). The Hamiltonian is
\begin{eqnarray}
\label{eq:ham}
 H &=& \sum_{k,i=\{L,R\}}\sum_{\sigma=1}^M \epsilon_k c_{ki\sigma}^{\dagger}
c_{ki\sigma} 
 + \frac{U}{2}(\hat{N}-N_g)^2
\nonumber\\
 & &
  +\;  \sum_{k,i=\{L,R\}}\sum_{\sigma=1}^M V_\sigma (c_{ki\sigma}^{\dagger}
d_{\sigma} + {\rm h.c.})
\end{eqnarray}
where the electron number operator for the impurity site is 
$\hat{N}\!=\!\sum_{\sigma=1}^M d_{\sigma}^{\dagger}d_{\sigma}$.
The energy in the bands is such that $-D \!\leq\epsilon_k\leq\! D$ where $D$ is
the half bandwidth, and we assume a flat density of states, $\rho \!=\! 1/2D$.
The hybridization of the impurity to each lead is given by $V_\sigma$ which
yields a level width $\Gamma_\sigma \!=\!
\Gamma_{L,\sigma}\!+\!\Gamma_{R,\sigma}$ with $\Gamma_{L,\sigma} \!=\!
\Gamma_{R,\sigma} \!=\! \pi V_\sigma^2 \rho$. In terms of the gate voltage
$N_g$, the energy level of the dot is explicitly given by $\epsilon_d \!=\! U(1
\!-\! 2N_g)/2$. Finally, in the absence of any orbital degeneracy, the
degeneracy of the $d$ level is simply given by spin, $\sigma\!=\uparrow$ or
$\downarrow$. 

\begin{figure}[t]
\centering
\includegraphics[width=2.7in]{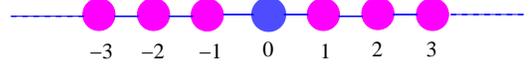}
\caption{(color online) The 1D infinite tight-binding chain, where the $0^{th}$
site is the impurity site (quantum dot).
}
\label{fig:chain}
\end{figure}

{\em Method---}A new basis for the two non-interacting bands can be
independently constructed by starting from the localized impurity state. In this
way the model is mapped to a one-dimensional (1D) infinite tight-binding chain
\cite{WilsonRMP75}, as shown in Fig.\,\ref{fig:chain}. We use a large chain
($\sim\! 10^6$ sites) in order that its finite size is irrelevant for the
physics of interest. Then, in order to make the computation time manageable,
logarithmic blocking of the energy levels \cite{WilsonRMP75} is used to reduce
the number of effective sites. In this work, the logarithmic blocking factor is
$\Lambda\!=\!2.5$ (the number of effective sites is $\sim\!61$). We use a form
of blocking \cite{Campo05} which avoids $\Lambda$-dependent corrections
\cite{Krishnamurthy80} to the low-energy scales [i.e.\ $\Tk(\Lambda)$]. We solve
the resulting problem using the world-line quantum Monte Carlo method with a
directed-loop cluster algorithm \cite{Syljuasen02,Yoo05}. The Trotter number $N$
is choosen such that $\varepsilon=\beta /N \!\simeq\! 0.1 / D$.

%The most important physical quantity with regard to quantum dot Kondo 
% physics experiments is the conductance at zero frequency. Extracting this 
% quantity from QMC calculations is problematic since world-line QMC operates 
% in imaginary time and the extrapolation or continuation of imaginary time 
% quantities to real time (or frequency) is known to be very 
% sensitive \cite{JarrellPhysRep96}. 

To find the conductance, we proceed following the method of Sylju\aa{}sen in
Ref.\,\onlinecite{Syljuasen07} which is itself closely related to several other
approaches \cite{IzumidaQMC99,Louis03,BokesPRB04,BokesJCP09}.
% NEED TO MENTION HAN SOMEWHERE??
The conductance at the (imaginary) Matsubara frequencies, $g(i\omega_n)$ with
$\omega_n\!=\!2\pi n T$, is related in linear response to the current-current
correlation function in the usual way. For a one-dimensional system with open
boundary conditions, current continuity can be used
\cite{Louis03,BokesPRB04,Syljuasen07} to express $g(i\omega_n)$ in terms of
charge correlations (polarizability), 
\begin{equation}\label{eq:giwn}
 g(i \omega_n) = \frac{\omega_n}{\hbar} \int_{0}^{\beta} d\tau \cos(\omega_n
\tau)\langle P_x(\tau) P_y(0) \rangle
\end{equation}
where $P_y$ is the sum of the electron charge density operators to the right of
$y$, $P_y \equiv \sum_{y^{'}\geq y}\hat{n}_{y^{'}}$. Thus the time derivative of
$\langle P_y \rangle$ is the current through the bond between sites $y-1$ and
$y$. We calculate $g(i\omega_n)$ for $n>0$ from the world-line QMC data in this
way. Not all combinations of $x$ and $y$ can be used in Eq.\,(\ref{eq:giwn})
because the system is not a physical chain but only effectively mapped to a
chain. Notice that the current through the four bonds closest to the impurity
site (labeled $0$) correspond to the physical current.
%(i.e.\ between sites -2 and -1, -1 and 0, 0 and 1, and 1 and 2). 
Therefore, $x$ and $y$ must be chosen from among $\{-1, 0, 1, 2\}$. In addition,
left-right symmetry reduces the number of independent combinations. In our
calculation, we choose three cases for $x$ and $y$: $(0,1)$, $(0,0)$, and
$(-1,0)$.

The linear conductance $G$ is obtained by extrapolating $g(i\omega_n)$ to zero
frequency, $G\!=\!\lim_{\omega_n\rightarrow 0}g(i\omega_n)$. We carry out this
extrapolation as follows. First, we try to fit the data at the four or five
lowest Matsubara frequencies [$g(i\omega_n)$ for $n\!=\!1,...,4$ or $1,...,5$]
to a linear or quadratic polynomial. If this method yields a good fit, we simply
extrapolate the data by using the polynomial. If neither polynomial fit is good,
the data at the first 14 lowest Matsubara frequencies are fit by using a series
of rational polynomial functions of different degree $[p/q]$  (e.g. $p$ for the
numerator, $q$ for the denominator, $p\!=\!1$ for a constant, $p\!=\!2$ for
linear function, etc.) as described in Ref.\,\onlinecite{Syljuasen07}. We use
all $p$ and $q$ such that $5\leq p+q\leq 10$ and $p,q\geq 2$ but exclude cases
in which spurious poles appears. The final extrapolated value is the average of
the results for these different forms, and the error bar at zero frequency is
the maximum spread, which is larger than the error bar of any single $[p/q]$
extrapolation. To justify this method, we check that three conditions are met.
(i)~The data for all the combinations of $x$ and $y$ must extrapolate to nearly
the same value (the current through different bonds at non-zero frequency can be
different, but current continuity requires that at zero frequency the current
through all bonds be the same). (ii)~The data should fit well to most of the
functional forms of degree $[p/q]$ (we cannot exclude too many cases).
(iii)~Finally, the conductance should have a small error bar (a large error bar
shows that the extrapolation is model dependent).

\begin{figure}[t]
\centering
\includegraphics[width=3.2in]{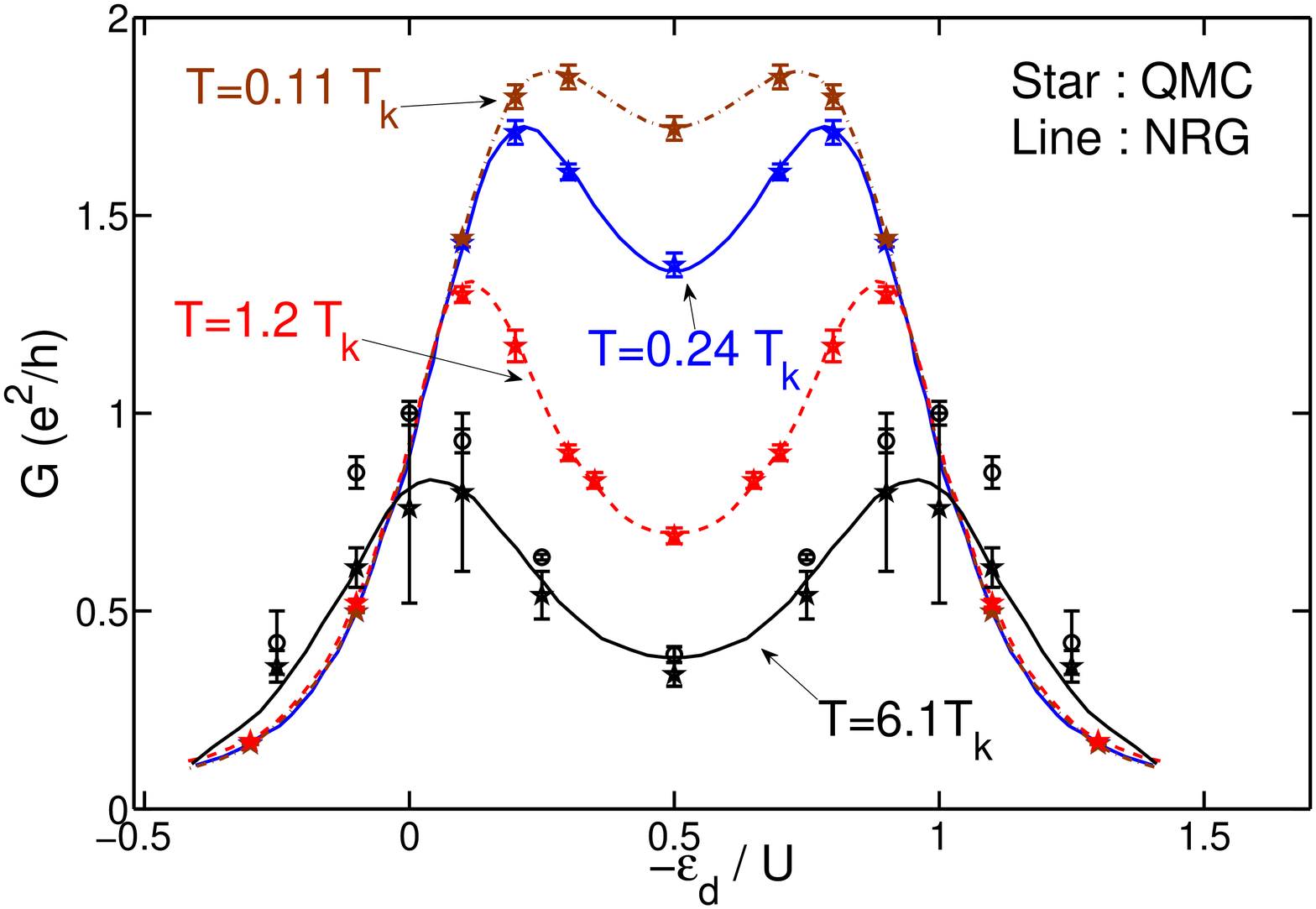}
\caption{(color online) Conductance through a single-level Anderson model
without orbital degeneracy as a function of gate voltage: QMC result (symbols)
compared to NRG calculation \cite{Costi01} (lines). 
Data for four temperatures are shown: $T\simeq 0.11 \,\Tk$ (brown, dot-dashed,
$\beta=98.3$), $T\simeq 0.24 \,\Tk$ (blue, upper solid, $\beta\!=\!43.7$),
$T\simeq 1.2 \,\Tk$ (red, dashed, $\beta\!=\!8.6$), and $T\simeq 6.1 \,\Tk$ (black,
lower solid, $\beta\!=\!1.7$). For $T\simeq 6.1 \,\Tk$, the black stars are for
$(x,y)\!=\!(0,1)$ while the black circles are for $(0,0)$. $\Tk$ denotes the
Kondo temperature found by NRG \cite{Costi01} at the particle-hole symmetric
point ($-\epsilon_d/U \!=\! 0.5$). Note the high accuracy of the QMC result as
long as $T \!\lesssim\! \Tk$.
}
\label{fig:condsu2}
\end{figure}

{\em Conductance without orbital degeneracy---}We first consider the standard
single-level Anderson model, $M\!=\!2$ in Eq.\,(\ref{eq:ham}). We compare the
conductance obtained by our QMC calculation to that from the numerical
renormalization group (NRG) calculation of Ref.\,\onlinecite{Costi01} [see their
Fig.~2(a)]. The parameters are $D\!=\!100$, $\Gamma\!=\!1.0$, and $U\!=\!3\pi$.
%($D\!=\!1$, $\Gamma\!=\!0.001$ and $U\!=\!3\pi\Gamma$ 
%in Ref.\,\onlinecite{Costi01}).
The NRG value \cite{Costi01} for the Kondo temperature at the particle-hole 
symmetry point ($-\epsilon/U\!=\!0.5$), which we denote $\Tk$ throughout, is 
$\Tk \!\simeq\! 0.1$.
%We set $\beta=1/T$. 

Fig.\,\ref{fig:condsu2} compares our calculation of the conductance as a
function of gate voltage to the NRG results \cite{Costi01} for several
temperatures. \textit{The QMC results are in excellent agreement with the NRG
results for $T \!\leq\! \Tk$ for all values of the gate voltage---that is, in
both the mixed valance and Kondo regimes.} For $T$ slightly larger than $\Tk$,
agreement is good; in contrast, note that there is a substantial error in the
extrapolated conductance value for larger $T$.

\begin{figure}[t]
\centering
\includegraphics[width=3.3in]{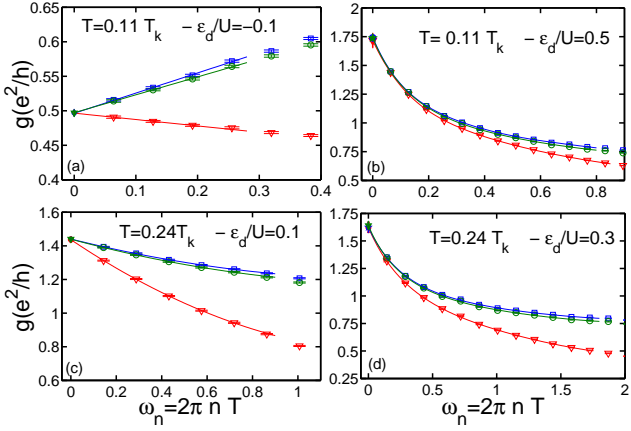}
\caption{(color online) Conductance at Matsubara frequencies at low temperature
(symbols) for the single-level Anderson model without orbital degeneracy and the
corresponding fits used to extrapolate to zero frequency (lines). The values of
$T$ and $-\epsilon/U$ are (a) $0.11 \,\Tk$, $-0.1$; (b) $0.11 \,\Tk$, $0.5$; (c)
$0.24 \,\Tk$, $0.1$; (d) $0.24 \,\Tk$, $0.3$. Points for three choices of
$(x,y)$ are shown: $(0,1)$ red triangles, $(0,0)$ blue squares, and $(-1,0)$
green circles. A good quality extrapolation is obtained in all cases.
}
\label{fig:smallTsu2}
\end{figure}

\begin{figure}[t]
\centering 
\includegraphics[width=3.2in]{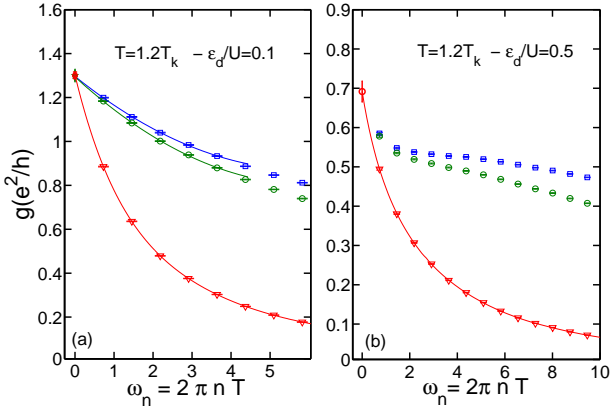}
\caption{(color online) Conductance at Matsubara frequencies for $T\!\simeq\!
1.2 \,\Tk$ in the absence of orbital degeneracy (symbols) and the corresponding
fits used to extrapolate to zero frequency (lines). The values of  $-\epsilon/U$
are (a) $0.1$ (mixed-valence) and (b) $0.5$ (Kondo regime). Points for three
choices of $(x,y)$ are shown: $(0,1)$ red triangles, $(0,0)$ blue squares, and
$(-1,0)$ green circles. Extrapolation using $(x,y)\!=\! (0,1)$ is accurate.
}
\label{fig:nearTksu2}
\end{figure}

\begin{figure}[t]
\centering
\includegraphics[width=3.2in]{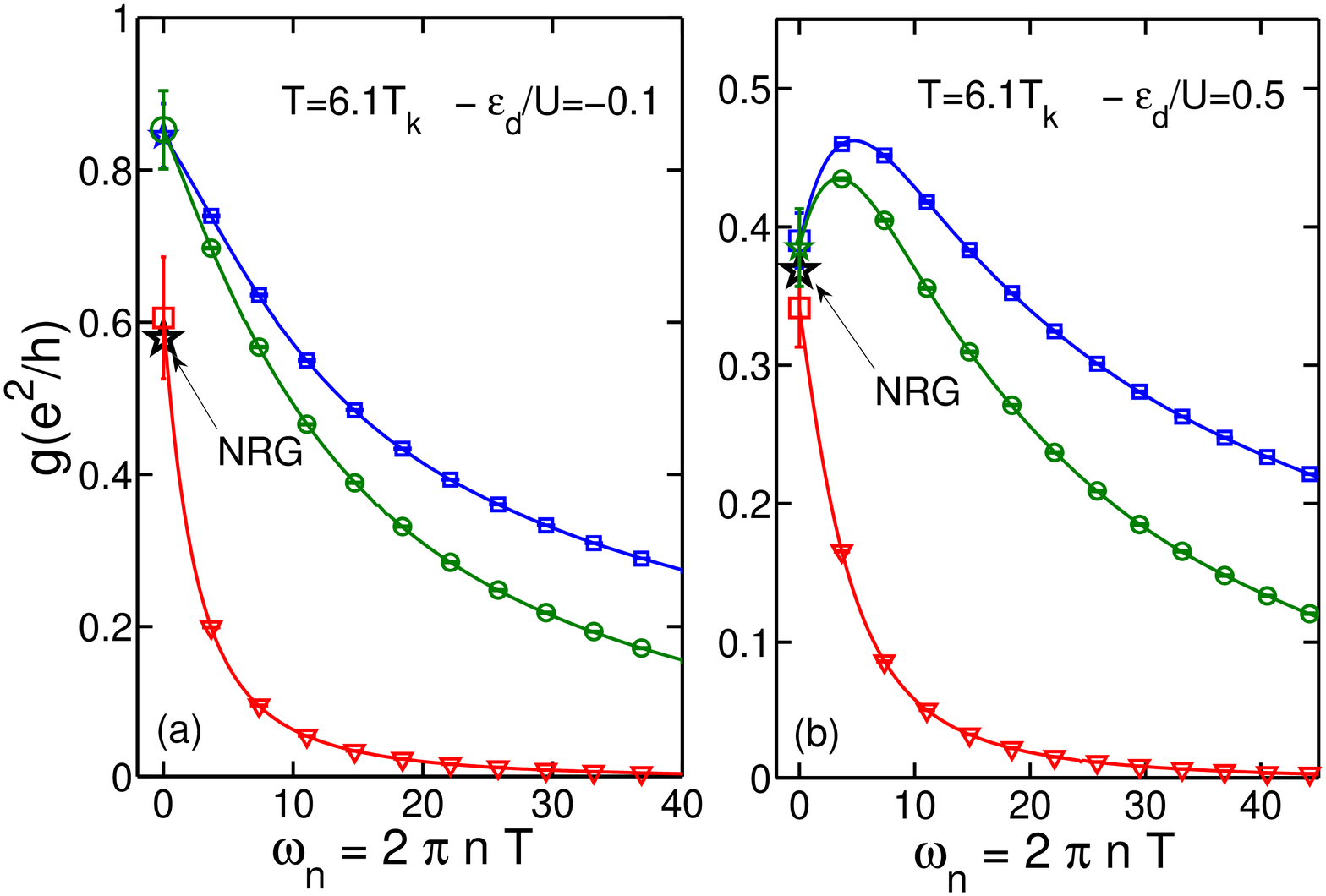}
\caption{(color online) Conductance at Matsubara frequencies for high
temperature, $T\!\simeq\! 6.1 \Tk$, in the absence of orbital degeneracy
(symbols) and the corresponding fits used to extrapolate to zero frequency
(lines). (a) $-\epsilon/U\!=\!-0.1$; (b) $-\epsilon/U\!=\!0.5$. Points for three
choices of $(x,y)$ are shown: $(0,1)$ red triangles, $(0,0)$ blue squares, and
$(-1,0)$ green circles. The black stars are the NRG data. The accuracy of the
extrapolation is poor in all of these cases.
}
\label{fig:largeTsu2}
\end{figure}

Some examples of the extrapolations used to obtain the conductance shown in
Fig.\,\ref{fig:condsu2} are given in
Figs.\,\ref{fig:smallTsu2}-\ref{fig:largeTsu2}, moving from lower to higher
temperature. Fig.\,\ref{fig:smallTsu2} shows four examples of the conductance at
imaginary frequency, $g(i\omega_n)$,  for $T \!<\! \Tk$. Examples of a linear
fit [panel (a)], a quadratic fit [panel (c)], and rational polynomial fits
[panels (b) and (d)] are shown. In the mixed valance regime, $-\epsilon_d/U
\!<\! 0.1$ or $>\! 0.9$, a linear or quadratic polynomial works well, and the
three curves for different $(x,y)$ all extrapolate to nearly the same value,
leading to a small error bar. In the Kondo regime, $0.1 \!<\! -\epsilon_d/U
\!<\!0.9$, the linear or quadratic polynomial does not fit well, but the QMC
data can be fit to a series of rational polynomials as discussed above. Almost
all values of $[p/q]$ work well, and the three sets of $g(i\omega_n)$ for
different $(x,y)$ extrapolate to nearly the same value, leading to a small error
bar. In this temperature regime, then, the extrapolation is straight forward and
the agreement with the NRG result is excellent.

For $T\!\sim\! \Tk$, two examples of the conductance function $g(i\omega_n)$ are
shown in Fig.\,\ref{fig:nearTksu2}. For the mixed valance regime
[Fig.\,\ref{fig:nearTksu2}(a)], a quadratic polynomial works well for
$(x,y)\!=\!(0,0)$ and $(-1,0)$, and rational polynomials are used for $(0,1)$.
All three combinations extrapolate to nearly the same value, so the result is
accurate. In the Kondo regime [panel (b),$-\epsilon_d/U\!=\!0.5$],
$g(i\omega_n)$ for $(0,1)$ can be fit with rational polynomials. However, for
both other cases, $(x,y)\!=\!(0,0)$ and $(-1,0)$, there is a small wiggle near
$\omega\!=\!2\pi \Tk$ in the imaginary frequency conductance function
$g(i\omega_n)$, showing that there is important structure below that frequency.
Since there is only one data point below $\omega\!=\!2\pi \Tk$, the
extrapolation is unreliable. Thus, we do not use the data when structure appears
at a frequency below which there are only a few data points. The conductance in
the Kondo regime for this temperature is based only on $(x,y)\!=\!(0,1)$;
nonetheless, the agreement with the NRG result is good.

Finally, for $T\!>\! \Tk$ (Fig.\,\ref{fig:largeTsu2}), the functions
$g(i\omega_n)$ for the three combinations of $(x,y)$ do not extrapolate to the
same zero-frequency value. Notice also that the conductance obtained in the
mixed-valence regime (the gate voltage at which the conductance peaks for this
temperature) has a large error bar. For the cases $(x,y)\!=\!(0,0)$ and
$(-1,0)$, the QMC data can be fit with a rational polynomial, but the
extrapolated result disagrees substantially with NRG. For the case $(0,1)$, the
average value of $G$ from QMC roughly follows the NRG result
(Fig.\,\ref{fig:condsu2}), but the large error bar in most cases indicates that
the result has little meaning. Thus, the QMC extrapolation method is unreliable
for $T$ substantially larger than $\Tk$.

\begin{figure}[t]
\includegraphics[width=3.2in]{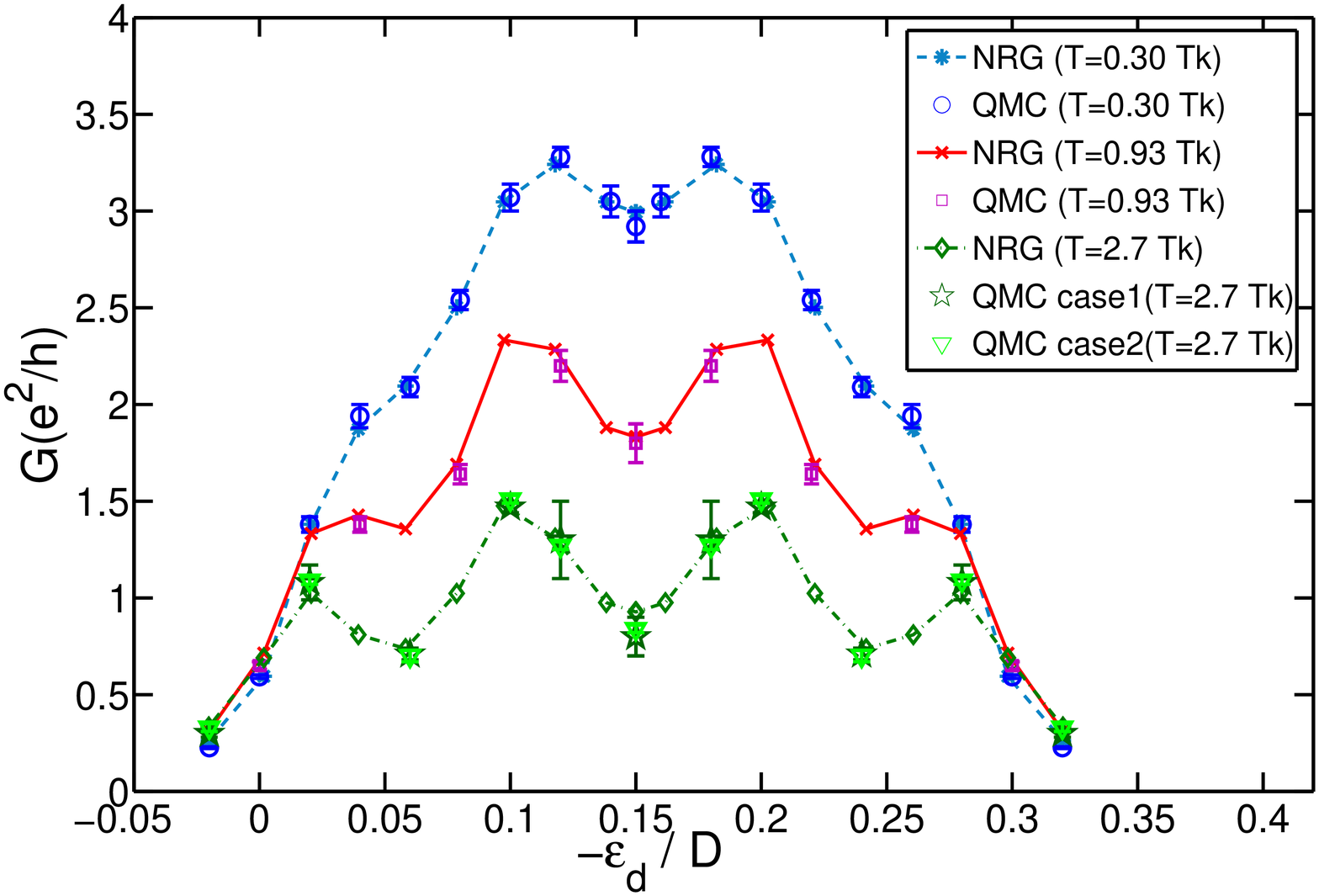}
\caption{(color online) Conductance in four-fold degenerate model as a function
of gate voltage: QMC results (symbols) compared with NRG calculations
\cite{Izumida98} (lines with symbols). 
Results for three temperatures are shown: $T\!\simeq\, 0.30 \,\Tk$ (blue circles
and solid line, $\beta\!=\!79.4$), $0.93 \,\Tk$ (red squares and dashed line,
$\beta\!=\!25.6$), and $2.7 \,\Tk$ (green triangles or stars and dotted line,
$\beta\!=\!8.77$).  
For the highest temperature ($T\simeq 2.7 \,\Tk$), the QMC data labeled case 1 are
based on $(x,y) \!=\! (0,0)$ and $(-1,0)$ while those for case 2 use $(0,1)$. 
$\Tk$ here denotes the Kondo temperature found by NRG \cite{Izumida98} at the 
particle-hole symmetric point.
The good agreement of the QMC data with the NRG results illustrates the value of
the QMC approach, though note the growing error bar when $T \agt \Tk$.
}
\label{fig:condsu4}
\end{figure}

{\em Conductance with orbital degeneracy---}We now turn to considering an
Anderson model in which all the states, both those in the dot and in the leads,
have an orbital degeneracy in addition to spin degeneracy: $M\!=\!4$ in
Eq.\,(\ref{eq:ham}). This situation arises, for instance, in carbon nanotube
quantum dots connected to carbon nanotube leads \cite{Jarillo05,
MakarovskiPRB07, MakarovskiPRL07, Choi05}. To assess the quality of our QMC
results, we compare with the NRG results of Ref.\,\onlinecite{Izumida98} (see
their Fig.\,16). The parameters we use are 
$D\!=\!30$, $U \!=\! 0.1 \,D \!=\! 3$, $\Gamma_{1,2} \!=\! 0.003\,\pi D$, and 
$\Gamma_{3,4} \!=\! 0.002\,\pi D$. 
% (D=1, $U = 0.1 D$, $\Gamma_e = 0.003\pi D$, and  $\Gamma_o = 0.002\pi D$ 
% in Ref.~\cite{Izumida98}). 
At the particle-hole symmetric point where the Kondo temperature is a minimum,
the NRG estimation \cite{Izumida98} for $\Tk$ yields $\Tk \simeq 0.0042$..

Fig.\,\ref{fig:condsu4} compares our calculation of the conductance as a
function of gate voltage to the NRG\cite{Izumida98} results. For $T\!\leq\!
\Tk$, the QMC and NRG results are in very good agreement throughout both the
mixed valence and Kondo regimes. For $T \!>\! \Tk$ ($T\simeq 2.7 \,\Tk$), the QMC
conductance roughly follows the NRG result but does not accurately agree with
it. In addition, a large error bar is encountered at the highest temperature,
showing that, as in the doubly degenerate case, the extrapolation is not
reliable for these temperatures.

\begin{figure}[b]
\centering
\includegraphics[width=3.3in]{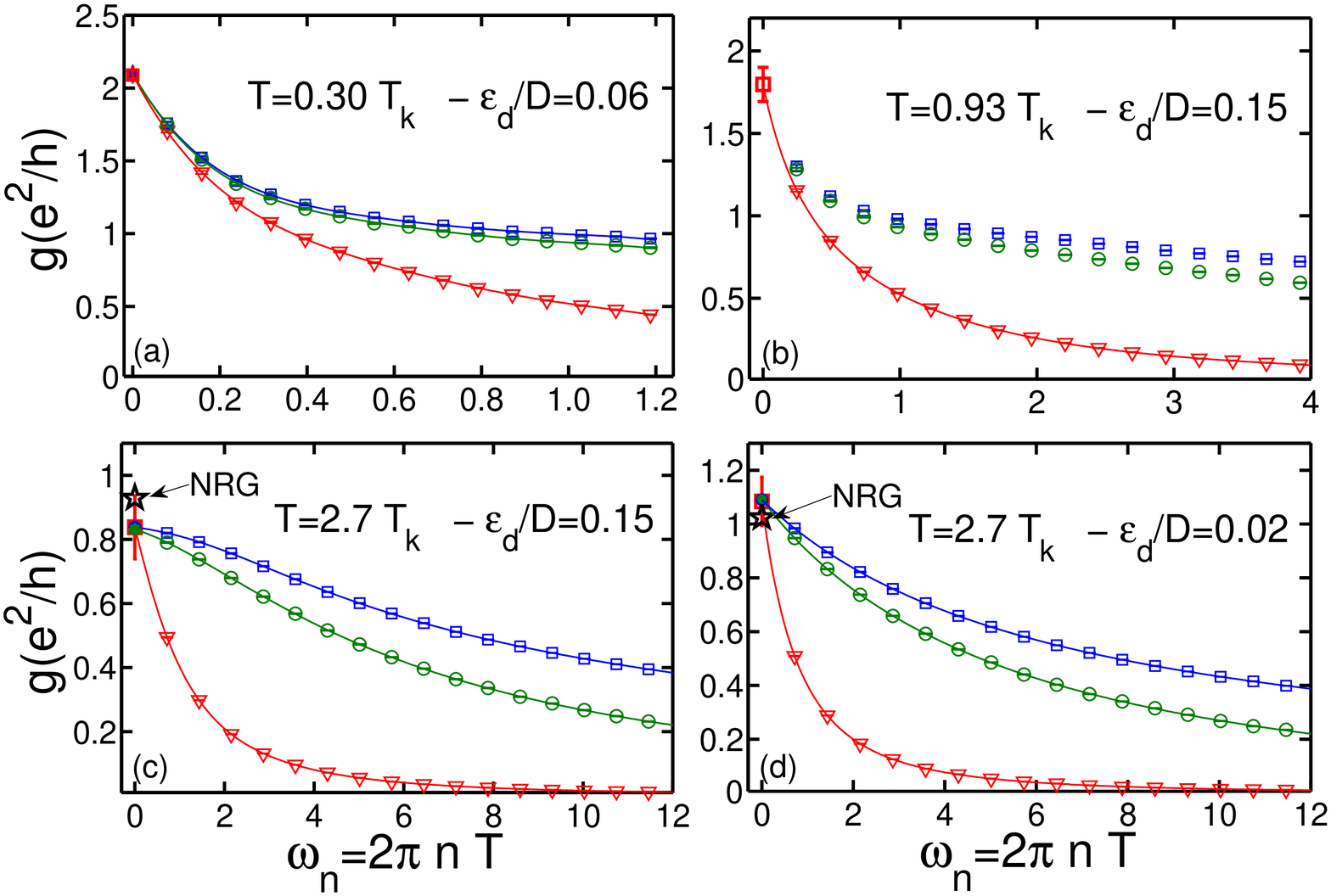}
\caption{(color online) Imaginary frequency conductance function for single
impurity Anderson model with orbital degeneracy ($M\!=\!4$). The values of $T$
and $-\epsilon/U$ are (a) $0.30 \,\Tk$, $0.06$, (b) $0.93 \,\Tk$, $0.15$ (near
particle-hole symmetry), (c) $2.7 \,\Tk$, $0.15$, (d) $2.7 \,\Tk$, $0.02$. Points
for three choices of $(x,y)$ are shown: $(0,1)$ red triangles, $(0,0)$ blue
squares, and $(-1,0)$ green circles. The black stars are the NRG data. The
extrapolation is successful at low temperature but becomes increasingly
problematic at higher temperature, $T \!>\! \Tk$.
}
\label{fig:allTsu4}
\end{figure}

Four examples of the extrapolation from the imaginary frequency conductance
function, $g(i\omega_n)$, are shown in Fig.\,\ref{fig:allTsu4}. At low
temperature, panel (a), the extrapolation is good and consistent for all three
values of $(x,y)$ using the rational polynomial fit. Near the particle-hole
symmetry point and for $T\!\simeq\! \Tk$ [panel (b)], the case with $(x,y) \!=\!
(0,1)$ fits nicely to a rational polynomial and the extrapolated value agrees
with NRG. For the other two curves ($(x,y) \!=\! (0,0)$ and $(-1,0)$), a small
wiggle appears near $\omega\!\simeq\! 2\pi \Tk$, as in the case without orbital
degeneracy (Fig.\,\ref{fig:nearTksu2}), making extrapolation difficult. For
larger $T$, panels (c) and (d), although the QMC data for two cases ($(x,y)
\!=\! (0,0)$ and $(-1,0)$) can be fit to rational polynomials and yield an
estimated conductance with small error bar, the value does not agree accurately
with the NRG result. The $(x,y) \!=\! (0,1)$ yields a large estimated error.
Therefore, as we saw in the case without orbital degeneracy, when the
temperature become large, the QMC method becomes inaccurate.

In summary, we developed and tested a method to obtain the linear conductance by
extrapolating from QMC data. By studying two cases for which NRG results exist
in the literature \cite{Costi01, Izumida98}, we demonstrated the accuracy of the
extrapolation technique as long as the temperature is not too high, $T \!\alt\!
\Tk$ (where $\Tk$ denotes the Kondo temperature at the particle-hole symmetric point). 
We expect that this technique will be useful for finding the conductance
of more complex quantum dot and/or impurity systems, such as three and four
quantum dot structures \cite{LindelofFujisawa08,Sachrajda09}.

We thank P. Bokes and J. Shumway for helpful discussions. This work was
supported in part by the U.S.\,NSF Grant No.\,DMR-0506953.

\bibliography{cond_qmc}

\end{document}